\documentstyle[lcws2002]{article}
\input{psfig}

\def\be{\begin{equation}}
\def\ee{\end{equation}}
\def\bea{\begin{eqnarray}}
\def\eea{\end{eqnarray}}
\def\Re{{\cal R \mskip-4mu \lower.1ex \hbox{\it e}\,}}
\def\Im{{\cal I \mskip-5mu \lower.1ex \hbox{\it m}\,}}
\def\ie{{\it i.e.}}
\def\eg{{\it e.g.}}
\def\etc{{\it etc}}
\def\etal{{\it et al.}}

\def\sub#1{_{\lower.25ex\hbox{$\scriptstyle#1$}}}
\def\tev{\,{\ifmmode\mathrm {TeV}\else TeV\fi}}
\def\gev{\,{\ifmmode\mathrm {GeV}\else GeV\fi}}
\def\mev{\,{\ifmmode\mathrm {MeV}\else MeV\fi}}
\def\mpl{\ifmmode M_{pl}\else $M_{pl}$\fi}
\def\to{\rightarrow}

\def\subw{_{\rm w}}
\def\mh{\ifmmode m\sbl H \else $m\sbl H$\fi}
\def\mch{\ifmmode m_{H^\pm} \else $m_{H^\pm}$\fi}
\def\mt{\ifmmode m_t\else $m_t$\fi}
\def\mc{\ifmmode m_c\else $m_c$\fi}
\def\mz{\ifmmode M_Z\else $M_Z$\fi}
\def\mw{\ifmmode M_W\else $M_W$\fi}
\def\mws{\ifmmode M_W^2 \else $M_W^2$\fi}
\def\mhs{\ifmmode m_H^2 \else $m_H^2$\fi}   
\def\mzs{\ifmmode M_Z^2 \else $M_Z^2$\fi}
\def\mts{\ifmmode m_t^2 \else $m_t^2$\fi}
\def\mcs{\ifmmode m_c^2 \else $m_c^2$\fi}
\def\mchs{\ifmmode m_{H^\pm}^2 \else $m_{H^\pm}^2$\fi}
\def\ztwo{\ifmmode Z_2\else $Z_2$\fi}
\def\zone{\ifmmode Z_1\else $Z_1$\fi}
\def\mtwo{\ifmmode M_2\else $M_2$\fi}
\def\mone{\ifmmode M_1\else $M_1$\fi}
\def\tb{\ifmmode \tan\beta \else $\tan\beta$\fi}
\def\xw{\ifmmode x\subw\else $x\subw$\fi}
\def\ch{\ifmmode H^\pm \else $H^\pm$\fi}
\def\lum{\ifmmode {\cal L}\else ${\cal L}$\fi}
\def\inpb{\,{\ifmmode {\mathrm {pb}}^{-1}\else ${\mathrm {pb}}^{-1}$\fi}}
\def\infb{\,{\ifmmode {\mathrm {fb}}^{-1}\else ${\mathrm {fb}}^{-1}$\fi}}
\def\epem{\ifmmode e^+e^-\else $e^+e^-$\fi}
\def\ppb{\ifmmode \bar pp\else $\bar pp$\fi}
\def\bsg{\ifmmode B\to X_s\gamma\else $B\to X_s\gamma$\fi}
\def\bsll{\ifmmode B\to X_s\ell^+\ell^-\else $B\to X_s\ell^+\ell^-$\fi}
\def\bstt{\ifmmode B\to X_s\tau^+\tau^-\else $B\to X_s\tau^+\tau^-$\fi}
\def\lamt{\ifmmode \tilde\lambda\else $\tilde\lambda$\fi}
\def\shat{\ifmmode \hat s\else $\hat s$\fi}
\def\that{\ifmmode \hat t\else $\hat t$\fi}
\def\uhat{\ifmmode \hat u\else $\hat u$\fi}

\newskip\zatskip \zatskip=0pt plus0pt minus0pt

\begin{document}

\rightline{\vbox{\halign{&#\hfil\cr
SLAC-PUB-9489\cr
September 2002\cr}}}
\vspace{0.8in}

\title{UNIQUE GRAVITON EXCHANGE SIGNATURES AT LINEAR COLLIDERS}

\author{ THOMAS G. RIZZO 
\footnote{Work supported by the Department of Energy, Contract 
DE-AC03-76SF00515}}

\address{Stanford Linear Accelerator Center,
Stanford,\\ CA 94309, USA}

\maketitle\abstracts{
Many types of new physics can lead to contact interaction-like modifications 
in $e^+e^-$ processes below direct production threshold. We examine the 
possibility of uniquely identifying the effects of graviton exchange from 
amongst this large set of models by using the moments of the angular 
distribution of the final state particles. In the case of  
$e^+e^-\to f\bar f(W^+W^-)$ we demonstrate that this technique allows 
for the unique identification of the graviton exchange signature at the 
$5\sigma$ level for mass scales as high as $6(2.5)\sqrt s$.}

It is generally expected that new physics beyond the Standard Model(SM) will 
manifest itself at future colliders that probe the TeV scale. 
This new physics(NP) may appear either directly, 
as in the case of new particle production, or indirectly through 
deviations from the predictions of the SM. 
Perhaps the most well known example of this indirect scenario 
would be the observation of deviations 
in, \eg, various $e^+e^-$ cross sections due to apparent contact 
interactions{\cite {big}}. There are many very different NP scenarios 
that predict new particle exchanges which lead to contact interactions 
below direct production threshold; 
a partial list of known candidates is: a $Z'$,  
leptoquarks, $R$-parity violating sneutrinos($\tilde \nu$), bileptons, 
graviton Kaluza-Klein(KK) towers, gauge boson KK towers, and even string 
excitations. 
If contact interaction effects are observed one can always try to fit the 
shifts in the observables to each one of the set of known theories and see 
which gives the best fit. 
Alternatively, one can devise a technique which will rather quickly divide the 
full set of all possible models into distinct subclasses. 
In this paper we propose such a technique that makes use of the 
specific modifications in angular distributions 
induced by new exchanges. This method 
offers a way to uniquely identify graviton KK tower exchange 
(or any possible spin-0 exchange).

Let us consider the normalized cross section for  
$e^+e^- \to f\bar f$ in the SM assuming $m_f=0$ and $f\neq e$ 
for simplicity. This can be written as 
$\sigma^{-1} d\sigma/dz={3\over {8}}(1+z^2)+A_{FB}(s)z$ 
where $z=\cos \theta$ and $A_{FB}(s)$ is the Forward-Backward Asymmetry which 
depends upon the electroweak quantum numbers of the fermion, $f$, as well as 
the center of mass energy of the collision, $\sqrt s$. This structure is 
particularly interesting in that it is equally valid for a wide variety of 
New Physics models: composite-like contact interactions, $Z'$ models, 
TeV-scale KK gauge bosons, \etc.  
In fact, when this expression holds the {\it only} deviation from the 
SM for any of these models will be through the variations in  
the value of $A_{FB}(s)$ since we have chosen to normalize the cross section. 

Now let us consider taking moments of the normalized cross section above with 
respect to the Legendre Polynomials, $P_n(z)$. This can be done easily by 
re-writing the above as 
$\sigma^{-1} d\sigma/dz=1/2 P_0+1/4 P_2+A_{FB}(s)P_1$ 
and recalling that the $P_n(z)$ are orthonormal. 
Denoting such moments as $<P_n>$, one finds that $<P_1>=2A_{FB}/3$, 
$<P_2>=1/10$ and $<P_{n>2}>=0$. In addition we also trivially obtain that 
$<P_0>=1$ since we have normalized the distribution so that this moment 
carries no new information. Thus, very naively, if 
we find that the $<P_{n>1}>$ are given by their 
SM values while $<P_1>$ differs 
from its corresponding SM value we could conclude that the NP is most likely 
one of those listed immediately 
above. If {\it both} $<P_{1,2}>$ differ from their SM expectations 
while the $<P_{n>2}>$ remain zero the source can only due to a 
spin-0 exchange in the $s-$channel. As we will see below only $s-$channel 
KK graviton exchange, since it is spin-2, leads to non-zero values of 
$<P_{3,4}>$ while the $<P_{n>4}>$ remain zero. Of course 
$<P_{1,2}>$ will also be different from their SM values in this case but 
as we have just observed this signal is {\it not} unique to gravity. 
This observation seems to yield a rather simple test for the 
exchange of graviton KK towers. It is important to note that 
we could not have performed this simple analysis 
for the case of Bhabha scattering, \ie, $e^+e^- 
\to e^+e^-$, as it involves both $s-$ and $t-$channel exchanges. 

The real world is not so simple as the idealized case 
we have just discussed for 
several reasons. First, we have assumed that we know the cross section 
precisely at all values of $z$, \ie, we have infinite statistics with no 
angular binning. 
Secondly, to use the orthonormality conditions we need to have complete 
angular coverage, \ie, no holes for the beam pipe, \etc. 
To get a feeling for how 
important these effects can be we consider dividing the distribution 
into a finite number of angular bins, $N_{bins}$, of common size 
$\Delta z=2/N_{bins}$. 
The results of this analysis are shown in Table 1; 
here we see that as the number 
of bins grows large we rapidly recover the continuum results discussed above. 
Of course in any realistic experimental situation, $N_{bins}$ remains finite
but a value of order 20 is reasonable as it strikes a respectable 
balance between the realistic 
demands of statistics and angular resolution. 
Now let us assume that $N_{bins}=20$ and examine the effects of the necessary 
cut at small angles due to the beam pipe, \etc. 
This is straightforward and leads to the 
results shown in Table 2 for various values of the small angle cut. 
Here we observe that the `background contamination' of the naive SM 
result increases quite rapidly as we make the angular cut stronger.

\begin{table}
\caption{Dependence on $N_{bins}$ for the first four moments of the 
normalized cross section for $e^+e^-\to f\bar f$ with $m_f=0$.   
Both $<P_{1,3}>$ are in units of $A_{FB}$.} 
\centering
\begin{tabular}{|c|c|c|c|c|} \hline\hline
$N_{bins}$&$<P_2>(10^{-2})$&$<P_4>(10^{-3})$&$<P_1>$  &$<P_3>(10^{-3})$  
\\ \hline \hline
   10   & 9.0040 &-26.7585    & 0.66000    &-23.1000      \\
   20   & 9.7503 & -6.8285    & 0.66500    & -5.8188      \\
   50   & 9.9600 & -1.0988    & 0.66640    & -0.9330      \\
$\infty$  & 10.0   &    0.0     &  2/3       &     0.0      \\ \hline\hline
\end{tabular}
\end{table}
\begin{table}
\caption{Dependence on the cut at small scattering angles in milliradions 
assuming $N_{bins}=20$ for the first four moments of the normalized cross 
section as in the previous Table. Both $<P_{1,3}>$ are in units of $A_{FB}$.} 
\centering
\begin{tabular}{|c|c|c|c|c|} \hline\hline
Cut(mr)&$<P_2>(10^{-2})$&$<P_4>(10^{-3})$&$<P_1>$  &$<P_3>(10^{-3})$  
\\ \hline \hline
   0    & 9.7503 & -6.8285    & 0.66500    & -5.8188      \\
   10   & 9.7428 & -6.8981    & 0.66490    & -5.9156      \\
   50   & 9.5652 & -8.5590    & 0.66251    & -8.2301      \\
  100   & 9.0159 & -13.616    & 0.65508    & -15.341      \\ \hline\hline
\end{tabular}
\end{table}

What this brief study indicates is that for a realistic detector 
the simple and naive expectations for the various moments will receive 
`backgrounds' that will need to be dealt with and subtracted from the real 
data to obtain information on the $<P_n>$. In the real world these 
backgrounds can be found through a detailed Monte 
Carlo study whose results will be influenced the detector geometry 
and by how well the properties of the detector are known. For our 
numerical analysis below we will follow a simpler approach by calculating the 
moments in the SM (after binning and cuts are applied) and then subtracting 
them from those obtained when the NP is present. 
Given the discussion above it is clear that we should begin by examining the 
process $e^+e^-\to f\bar f$. To be specific we concentrate on the 
model of Arkani-Hamed \etal, ADD, though our results are easily 
extended to the Randall-Sundrum model below   
graviton production threshold. 
The differential cross section for $e^+e^-\to f\bar f$, now including graviton 
tower exchange, for massless fermions can be written as   
\begin{eqnarray}
\label{dsdz}
{d\sigma\over dz} & = & N_c{\pi\alpha^2\over s}\left\{ \tilde P_{ij}
\left[A^e_{ij}A^f_{ij}(2P_0+P_2)/3+2B^e_{ij}B^f_{ij}P_1\right]\right.
\nonumber\\
& & -{\lambda s^2\over 2\pi\alpha \Lambda_H^4}\tilde P_i
\left[v^e_iv^f_i~(2P_3+3P_1)/5+a^e_ia^f_i P_2 \right] \\
& & \left. +{\lambda^2s^4\over 16\pi^2\alpha^2 \Lambda_H^8}\left[
(16P_4+5P_2+14P_0)/35\right]\right\} \,,
\nonumber
\end{eqnarray}
where the indices $i,j$ are sum over the $\gamma$ and $Z$ exchanges, 
$\tilde P_{ij}$ and 
$\tilde P_i$ are the usual dimensionless propagator factors, 
$A^f_{ij}=(v_i^fv_j^f
+a_i^fa_j^f)\,, B^f_{ij}=(v_i^fa_j^f+v_j^fa_i^f)\,,$ and $N_c$ 
represents the number of final state colors. $\Lambda_H$ is the cutoff 
scale employed by Hewett in evaluating the summation over the 
tower of KK 
propagators and $\lambda=1$ will be assumed. Our results will not 
depend upon this choice of sign. This cross section has an 
explicit dependence on the $P_{n>2}$ associated with the exchange 
of the graviton tower. Terms proportional to $P_{3}$ occur 
in the interference between the SM and gravitational contributions whereas the 
term proportional to $P_{4}$ occurs only in the pure gravity piece. This 
implies that for $\sqrt s <<\Lambda_H$ it will be $<P_{3}>$ which 
will show the largest shifts from the expectations of the SM. 
With the polarized beams that we expect to have available at a linear 
collider, a z-dependent Left-Right Asymmetry, $A_{LR}$, can also 
be formed which provides an additional observable. 

Our approach will be as follows: we consider two observables ($i$) the 
normalized unpolarized cross section and ($ii$) the normalized difference 
of the polarized cross sections $\sim (d\sigma_L-d\sigma_R)/dz$, which is 
essentially given by $A_{LR}$. We 
then calculate the first four non-trivial moments of these two observables 
for the $\mu,\tau, b,c$ and $t$ final states within the SM including the 
effects of Initial State Radiation(ISR). (Note that for the $t\bar t$ 
final state we need to include finite mass effects.) 
Here we will assume tagging 
efficiencies of $100\%$, $100\%$, $80\%$, $60\%$ and 
$60\%$, respectively, for the various final states and that $N_{bins}=20$ with 
$\theta_{cut}=$50mr. The resulting values for 
the $<P_n>$ as calculated in the SM will be called `background' values 
consistent with our discussion above. Next, we calculate the 
same moments in the ADD model by choosing a  
value for the parameter $\Lambda_H$. Combining both observables and summing 
over the various flavor final states we can form a $\chi^2$ from the 
deviation of the $<P_{3,4}>$ moments from their SM `background'  
values. For a fixed integrated 
luminosity this can be done using the statistical errors 
as well as the systematic errors associated with the precision expected on 
the luminosity and polarization measurements. (Here we will assume the values 
$\delta L/L=0.25\%$ and $\delta P/P=0.3\%$.) 
Next we vary the value of the scale $\Lambda_H$ until we obtain a 
$5\sigma$ deviation from the SM; we call this value of $\Lambda_H$ the 
Identification Reach as it is the maximum value for the scale at which we 
observe a $5\sigma$ deviation from the SM values of $<P_{3,4}>$ which we now 
know can only arise due to the effects of graviton exchange. Note that 
this value of the scale should {\it not} be confused with the Discovery Reach 
at which one observes an overall deviation from the SM. 
Although both the $<P_{1,2}>$ {\it also} deviate from their SM 
values these shifts cannot be directly attributed to a spin-2 exchange. (As 
noted previously, the shift in $<P_{1}>$ results for any of the 
NP models listed 
above whereas a shift in $<P_{2}>$ occurs whenever the new $s$-channel 
exchange is {\it not} spin-1, \eg, $\tilde \nu$ exchange.)
The results of these calculations are shown in Fig.1 for several values of 
$\sqrt s$ as a function of the integrated luminosity. Specifically, 
for a $\sqrt s=500$ GeV machine with 
an integrated luminosity of $1~ab^{-1}$ the ID reach with single(double) beam 
polarization is found to be 2.6(3.0) TeV, \ie, $(5-6)\sqrt s$. We remind the 
reader that the corresponding search reach for these luminosities is in 
range of $(9-10)\sqrt s$. Note that the ID reach we obtained by is a rather 
respectable fraction of the corresponding search reach.

\begin{figure}[htbp]
\centerline{
\psfig{figure=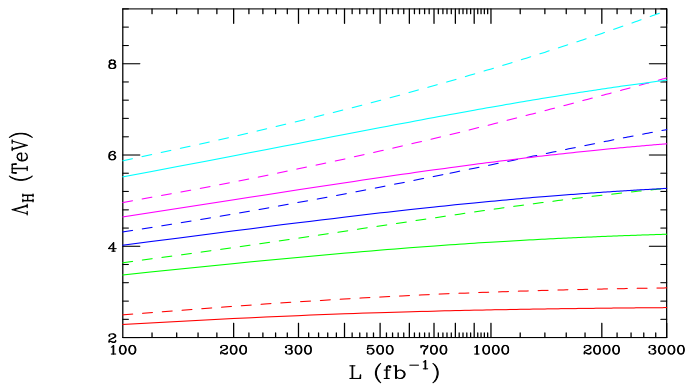,height=5cm,width=9cm,angle=90}}
\vspace*{0.1cm}
\caption{Identification reach in $\Lambda_H$ as a function of the integrated 
luminosity for the process $e^+e^-\to f\bar f$, with $f$ summed over 
$\mu,\tau,b,c$ and $t$. The solid(dashed) curves are for an $e^-$ polarization 
of $80\%$(together with a $e^+$ polarization of $60\%$). From bottom to top 
the pairs of curves are for $\sqrt s=0.5, 0.8,1, 1.2$ and $1.5$ TeV, 
respectively.}
\end{figure}

Other SM processes with large tree-level cross 
sections in which gravitons can be exchanged are $e^+e^- \to e^+e^-, 
\gamma\gamma, ZZ$, and $W^+W^-$ all of which involve $t-$ and/or 
$u-$channel exchanges. This would apparently disqualify them from further 
consideration. The $W^+W^-$ case is, however, special  because 
the $t-$channel $\nu$ exchange can be removed through the use of 
right-handed beam polarization. 
The remaining purely right-handed 
SM cross section is only quadratic in $z$ and this will not 
change if, \eg, $Z'$ or new $s-$channel scalar contributions are also present. 
The difficulty in this case is 
that the left-handed cross section is much larger than 
the right-handed one so that the possibility of `contamination' from the wrong 
polarization state is difficult to eradicate unless very good control over the 
beam polarization is maintained. Apart from the problem of isolating the 
purely right-handed part of the cross section  
we might conclude that the non-zero $<P_{3,4}>$ moments will be be a 
unique signature of graviton exchange for this process. Futhermore, since the 
pure gauge sector of the SM individually conserves $C$ and $P$, there are no 
terms in the cross section linear in $z$ and thus $<P_1>$ is expected to be 
zero in the SM and in many NP extensions. Such terms are, however, 
generated by KK graviton exchange so that a 
non-zero value of $<P_1>$ is also a gravity probe.

Apart from new particle exchanges there is another source of NP that 
can modify the right-handed $W$-pair cross section in a manner 
similar to gravity: anomalous gauge 
couplings(AGC). As is well known AGC can be $C$ and/or $P$ 
violating; one that violates both $C$ and $P$ but is $CP$ conserving can 
produce non-zero $<P_{1,3,4}>$ moments. Decomposing the $WWV$ 
($V=\gamma,Z$) vertex in the most general way 
yields 7 different anomalous couplings for each $V$ with the corresponding  
form factors denoted by $f_i^V$. (When weighted by the sum over 
the $\gamma$ and $Z$ propagators in $e^+e^-\to W^+W^-$ 
these form factors will be denoted by $F_i$, only two of 
which, $F_{1,3}$, are non-zero at the tree-level in the SM.) 
There is a single term in this general vertex expression with 
the correct $C$ and $P$ properties: that proportional to 
$f_5^V\epsilon^{\mu\alpha\beta\rho}(q^- -q^+)_\rho\epsilon_\alpha(W^-)
\epsilon_\beta(W^+)$ with $q^\pm$ the outgoing $W^\pm$ 
boson momenta, the $\epsilon$'s being their corresponding polarizations 
and $f_5^V$ being the relevant form factor. 
As we will see such a term will generate non-zero values for 
all of $<P_{1,3,4}>$. 
It appears that the possibility of non-zero AGC would 
contaminate our search for unique graviton exchange signatures. 
There is a way out of this dilemma; while gravity induces non-zero values 
for the $<P_{1,3,4}>$ from  
the angular distributions for $e^+e^-_R \to W^+W^-$ {\it independently}
of the final state $W$ polarizations, 
the $f_5^V \neq 0$ (\ie, $F_5$) couplings  
only contribute to the final state with mixed 
polarizations, \ie, transverse plus longitudinal, $TL+LT$. We recall that 
by measuring the angular distribution of the decaying $W$ relative to its 
direction of motion we can determine its state of polarization. 
Writing $d\sigma_R/dz \sim \Sigma_{TT}+\Sigma_{LL}+\Sigma_{TL+LT}$, 
one finds that in the SM both 
the $TT$ and $LL$ terms are proportional to $1-z^2$ while the $TL+LT$ term is 
proportional to $1+z^2$; no terms linear in $z$ are present. A non-zero 
$F_5$ induces additional terms in the case of the 
$TL+LT$ final state which now contains linear, cubic and quartic powers of 
$z$ similar to that generated by 
gravity. However, the $TT$ and $LL$ final states 
receive no such contributions. Thus 
observing non-zero values of $<P_{1,3,4}>$ (again, above backgrounds) for $W$ 
pairs in the $TT+LL$ final states produced by right-handed electrons 
{\it is} a signal for KK graviton exchange.  

\begin{figure}[htbp]
\centerline{
\psfig{figure=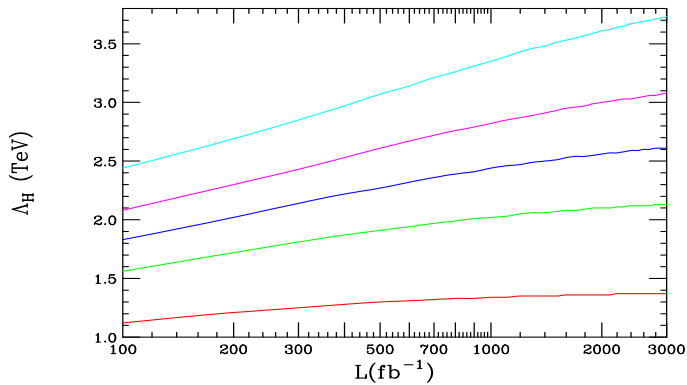,height=5cm,width=9cm,angle=90}}
\vspace*{0.1cm}
\caption{Same as in the previous figure but now for the $W^+W^-$ final state 
using the combined analysis discussed in the text.}
\end{figure}

Unfortunately we can never obtain purely RH beams at a LC. 
The only possibility is to measure the $W^+W^-$ cross section with two or 
more sets of different beam polarizations and then attempt to extract the 
purely right-handed piece from these measurements, again keeping only the 
$TT+LL$ contributions. In the case of two 
polarized beams this is perhaps best demonstrated by examining what happens 
when we combine two sets of data: one with $P(e^-)=-80\%$ and $P(e^+)=60\%$ 
and the other with both polarizations flipped. 
In comparison to the purely right-handed case, 
here we suffer from having to be able to very precisely subtract the 
additional large backgrounds arising from 
the left-handed parts of the cross section. 
In addition, both reduced statistics (since the luminosity is 
divided between both measurements) and the systematic errors associated with 
the polarization uncertainties will lead to further reductions in the 
anticipated identification reach. 
Fig.2 shows the results of this analysis. Here we see that the 
identification reach at large luminosities saturates due to the size of the 
systematic errors in extracting the right-handed piece of the cross section. 
The $5\sigma$ identification 
reach is found to be roughly $\sim 2.5\sqrt s$ for integrated luminosities 
of order $1~ab^{-1}$ which is far below that found for fermion pairs. 
It is unlikely that a more judicious choice of beam polarizations could 
drastically increase this reach. 

Many new physics scenarios predict the existence of contact interaction-like 
deviations from SM cross sections at high energy $e^+e^-$ colliders. 
We have found a technique that yields $5\sigma$ ID reaches for graviton 
exchange of 6(2.5)$\sqrt s$ for the processes $e^+e^- \to f\bar f(W^+W^-)$.


\section*{References}
\begin{thebibliography}{99}
%
\bibitem{big}
For details of the analysis and original references, see T.G. Rizzo, 
hep-ph/0208027.
\end{thebibliography}
\end{document}